\documentclass[9pt,twocolumn,twoside]{pnas-new}

\templatetype{pnasresearcharticle} 

\usepackage{ulem}

\title{Disordered biopolymer filament bundles: Topological defects and kinks}

\author[a]{Valentin M. Slepukhin}
\author[b]{Maximilian J. Grill}
\author[c,d]{Qingda Hu}
\author[c,d,e]{Elliot L. Botvinick}
\author[b]{Wolfgang A. Wall}
\author[a,f,g]{Alex J. Levine}

\affil[a]{Department of Physics and Astronomy, UCLA, Los Angeles California, 90095-1596, USA}
\affil[b]{Institute for Computational Mechanics, Technical University of Munich, Germany}
\affil[c]{Department of Biomedical Engineering, UCI, Irvine California,  CA 92697-2730, USA}
\affil[d]{Center for Complex Biological Systems, UCI, Irvine California, CA 92697-2280, USA}
\affil[e]{Beckman Laser Institute, UCI, Irvine California,  CA 92697-2730, USA}
\affil[f]{Department of Chemistry and Biochemistry, UCLA, Los Angeles California, 90095-1596, USA}
\affil[g]{Department of Biomathematics, UCLA, Los Angeles California, 90095-1596, USA}

\leadauthor{Slepukhin}

\significancestatement{A common structural motif for stress-bearing intracellular structures and tissues is 
a network of filament bundles.  When observed in optical microscopy, these bundles often show localized bends
-- kinks -- in their contour even though the stress-free state of their constituent filaments is straight. 
Using a combination of analytical mechanics and large-scale, finite-element Brownian dynamics simulations, 
we show that the optically observable kinks are related to topological defects in the interior, nanoscale
structure of the bundle.  Moreover, the kinks are more compliant to bending than the rest of the bundle. As
a result, defected bundle networks must contain a random distribution of soft hinges, which being the 
most compliant elastic elements, control the low energy excitations of these bundle networks.}

\authorcontributions{VMS and AJL performed the theoretical calculations.  MJG and WAW performed the simulations. 
QH and EB performed the experiments.}
\authordeclaration{We have no conflict of interest.}
\correspondingauthor{\textsuperscript{2}Valentin Slepukhin. E-mail: valentinslepukhin@physics.ucla.edu}

\keywords{Semiflexible filaments $|$ Bundles $|$ Topological defects $|$ Elasticity}

\begin{abstract}
Bundles of stiff filaments are ubiquitous in the living world, found both in the 
cytoskeleton and in the extracellular medium. These 
bundles are typically held together by smaller cross-linking molecules. We demonstrate 
analytically, numerically and experimentally that 
such bundles can be kinked, i.e., have localized regions of high curvature that are long-lived
metastable states. We propose three possible mechanisms of kink stabilization: 
a difference in trapped length of the filament segments  
between two cross links; a dislocation where 
the endpoint of a filament occurs within the bundle, and the braiding of the filaments 
in the bundle. At a high concentration of cross links, the last two effects lead to the 
topologically protected kinked states. Finally, we explore numerically and analytically
the transition of the metastable kinked state to the stable straight bundle.
\end{abstract}

\dates{This manuscript was compiled on \today}
\doi{\url{www.pnas.org/cgi/doi/10.1073/pnas.XXXXXXXXXX}}

\begin{document}

\maketitle
\thispagestyle{firststyle}
\ifthenelse{\boolean{shortarticle}}{\ifthenelse{\boolean{singlecolumn}}{\abscontentformatted}{\abscontent}}{}

\dropcap{S}emiflexible biopolymer filaments, i.e., stiff filaments whose thermal persistence 
length is comparable to their length, form most of the structural elements within cells and in the 
extracellular matrix surrounding them in tissues.   Common intracellular examples include the F-actin 
and intermediate filaments forming the cytoskeleton, while the extracellular matrix making up most 
tissues is composed of other stiff filamentous structures, such as 
collagen and elastin fibers.  The three-dimensional structure of these fiber networks is typically 
fixed by a variety of specific cross-linking proteins. On a smaller scale, these filaments often share a 
similar structural motif -- they form bundles of nearly aligned filaments, which are often 
densely cross linked along their contours.

While bundles might be regarded merely as new and thicker (thus stiffer) filaments, this 
analysis is inadequate in detail. For instance,
bundle bending mechanics can dramatically differ from those of a simple filament because, 
by having extra degrees of freedom associated with sliding one constituent filament 
relative to another within the bundle, the bundle acquires a length-dependent
effective bending modulus~\cite{bathe:06,Frey:07}.   These internal degrees of freedom 
also suggest that a nearly parallel group of filaments when quenched into a bundle by the 
addition of cross-linking agents may end up in one of many metastable states in which 
cross linking traps a defect, i.e., a long-lived structure distinct from elastic ground 
state of straight, parallel, and densely cross-linked filaments.  We focus on these 
defected, metastable states and their 
effect on the low-energy configurations of the bundle. Specifically, we show that 
there are three types of defects, two of which correspond to topological defects in the 
bundle's unstressed state -- braids and dislocations.  These and a third form of 
trapped length (loops) are all long-lived structures due to cross linking. 

As a result of these structural defects within the bundle, the elastic reference state is no 
longer straight, even though straight filament configurations are individually the lowest 
energy state of the constituent filaments.  Bundles containing these defects 
can minimize their elastic energy by taking on localized bends, which we call kinks.  The presence 
of kinks allows one to relate the micron-scale contour of kinked filament bundles to their 
nanoscale structure, specifically the presence of length-trapping defects.  We show that 
the combination of theory and simulation of defected bundles can account for the 
distribution of kinks we observe in experiment.  
Over long times, defects slowly anneal in bundles. This slow relaxation of the bundle's 
structure can be understood in terms of the diffusion and interaction of the defects on it. 
Specifically, defects leave the bundle either through diffusion 
off the bundle's ends, or by the annihilation of defects within it. 

Defected bundles not only explain the apparent kinks in collagen fibers, but also the 
presence of defects have implications for the collective elastic response of the bundle.  
In particular, we show that kinks are more bending compliant than undefected lengths of a 
bundle.  As a result, we hypothesize that the collective mechanics of a network of defected 
bundles depends on the number and position of these quenched defects, which act like soft hinges 
in a three-dimensional network of bundles that behave more like stiff beams.  

In our observations of collagen networks, we observe kinked bundles, whose 
contour we quantify by measuring their local curvature using light microscopy.  Due to 
their connection to the network, we cannot be certain that these kinks are not in some way 
related to elastic stress in the network.  To address this question, we used large-scale 
Brownian dynamics simulations to study kinking in quenched filaments with force- and 
torque-free boundary conditions, finding that quenched defects produce a statistical 
distribution of kinks similar to those observed in the experiment.  
Using the simulations, we are also able to measure the reduction of the bundle's local bending 
modulus at the location of the defects and observe the motion of the defects along the bundle.
Finally, we present theoretical calculations using a simple model of semiflexible filaments 
that demonstrate the relationship between defects and kinks in the bundle.  
Moreover, we analytically determine (and test via simulation) the time evolution of the
number of defects in a bundle as they slowly anneal through defect-defect
annihilation or by diffusion off the ends.

We first report our observations from light microscopy of kinks in collagen bundles and compare 
these kinks with those from numerical simulations. We then present a general discussion of the three
types of defects and demonstrate that the minimum energy state of the defected bundle can be kinked.
We explore defect dynamics, estimating the life time of a kink and the number of kinks in a 
bundle as a function of time, which we compare to simulation. To properly describe interaction 
of braiding type defects, we use the theory of the braid group; some relevant background is 
provided in the SI.

\section*{Results}
\label{sec:Results}
\subsection{Experiment}
\label{sec:Experiment}
\begin{figure*}[htpb]
\centering
\includegraphics[width=0.85\linewidth,trim=0 1.5cm 6cm 0, clip]{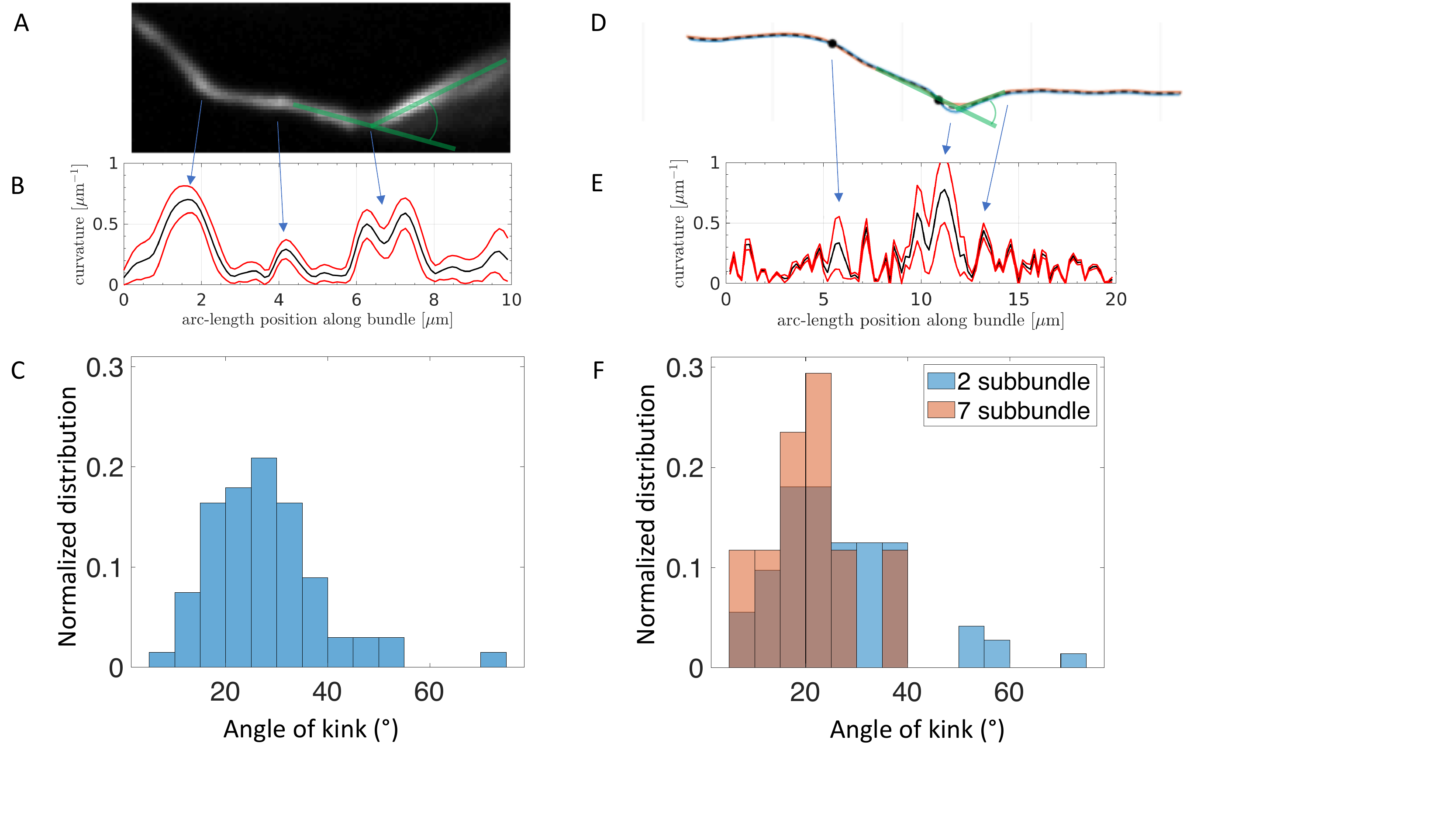}
\caption{Measurements of collagen bundles using fluorescence microscopy (left column) and simulations (right). 
A) Fluorescence image of a collagen bundle with noticeable kinks. Green traces show a measurement of 
the kink angle, given as angular deviation from straight. B) Mean curvature (black) of a 
collagen bundle over 50 images. Red lines show the spread of curvature in time (1 std). Blue 
arrows mark the locations of high curvature. C) Histogram of mean kink angles for all measured bundles ($n=74$). 
D) Simulation of bundles with reversible cross linkers showing the bundle centerline (black dashed line),
individual filaments (blue and red) and braids (black dots). E) Mean of the curvature over 100 
configurations (black). Red lines indicate one standard deviation. F) Kink angles from 
simulation for 2 or 7 filament bundles measured in the same way as in (C).}
\label{fig:kink_images_angles_and_curvature}
\end{figure*}

The nanoscale structure of collagen is quite complex~\cite{shoulders2009collagen,buehler2006nature}.
Small fibrils bind together to form larger fibrils, which, in turn, bind together to form fibers, which we observe in
light microscopy. Given that these fibers associate rapidly and strongly with local bonds, 
collagen fibers are a good place to look for quenched defects in bundles and kinks, if such sharp
bends of the bundle indeed result from those defects~\cite{kannus:2000}. In fact, kinked 
collagen bundles have been observed previously~\cite{franchi:2010} using electron microscopy. 
These observations leave the possibility that the kinks observed in a single snapshot of a dynamic,
flexible structure may be consistent with thermal undulations about a straight 
equilibrium state, rather than long-lived sharp bends ~\cite{wakuda2018native}.  
To address this question, we made multiple observations of collagen bundles in an aqueous 
environment to determine if the time-averaged state of the bundles includes kinks.

We reconstituted pepsin-extracted type I bovine collagen and fluorescently labelled and imaged individual bundles. In
Fig.~\ref{fig:kink_images_angles_and_curvature}A we show fifty superimposed images of a single
bundle (White on a black background) taken $0.5$s apart and showing three persistent kinks,
which confirms that they are indeed long-lived structures. 
Green lines indicate the measurement of a kink angle. 74 kink angle measurements from 43 bundles are 
summarized in Fig.~\ref{fig:kink_images_angles_and_curvature}C. The trace of the local 
curvature versus arc length along the bundle shown in
Fig.~\ref{fig:kink_images_angles_and_curvature}B quantifies the points of high 
persistent curvature as indicated by the blue arrows.  These local curvatures were computed by 
discretizing the contour using the intensity pixels in each image and computing the curvature from a cubic 
spline fit to these data. More details are given in SI Sec.~1B. 
Repeating this procedure for other bundles, we observed kinks and determined their mean kink 
angles by averaging again over up to fifty repeated measurements of each kink angle. 
They showed temporal fluctuations with a nonzero mean. We present the distribution 
of kink angles for 74 bundles in Fig.~\ref{fig:kink_images_angles_and_curvature}C. 
There were larger variations between kink angles measured across multiple bundles than in the 
thermal fluctuations of a given kinked bundle.  The distribution of 
these time-averaged kink angles has a mean at $26$ degrees and includes a range of 
typical angles between $7$ and $55$ 
degrees.  We observed one high-angle kink with a bend of $74$ degrees.

Many of the experimentally observed kinks appeared to be flexible. As typical example, the 
kink angle of the bundle shown in Fig.~S1 had a mean of 29 degrees, but fluctuated between 
21 and 38 degrees.  Because the bundle's ends were constrained by the network, we cannot use these 
thermal fluctuations
of the kink angle as a true measure of the kink's bending compliance. Another 
source of error is determining the location of bundle in the images, which is complicated 
by imaging resolution and noise, and by the limited plane of imaging (see Fig.~S2). 
In simulations, where we have control over the mechanical boundary conditions of the bundles and 
know the exact position of the bundles, we will return to this point.

\subsection{Numerical simulation}
\label{sec:Simulation}
To better explore both the nanoscale structure of the cross section of the kinked bundles and to
study the system with simpler, free 
boundary conditions, we turn to Brownian dynamics, finite-element simulations. 
\begin{figure*}[htpb]
  \centering
  \includegraphics[width=\linewidth]{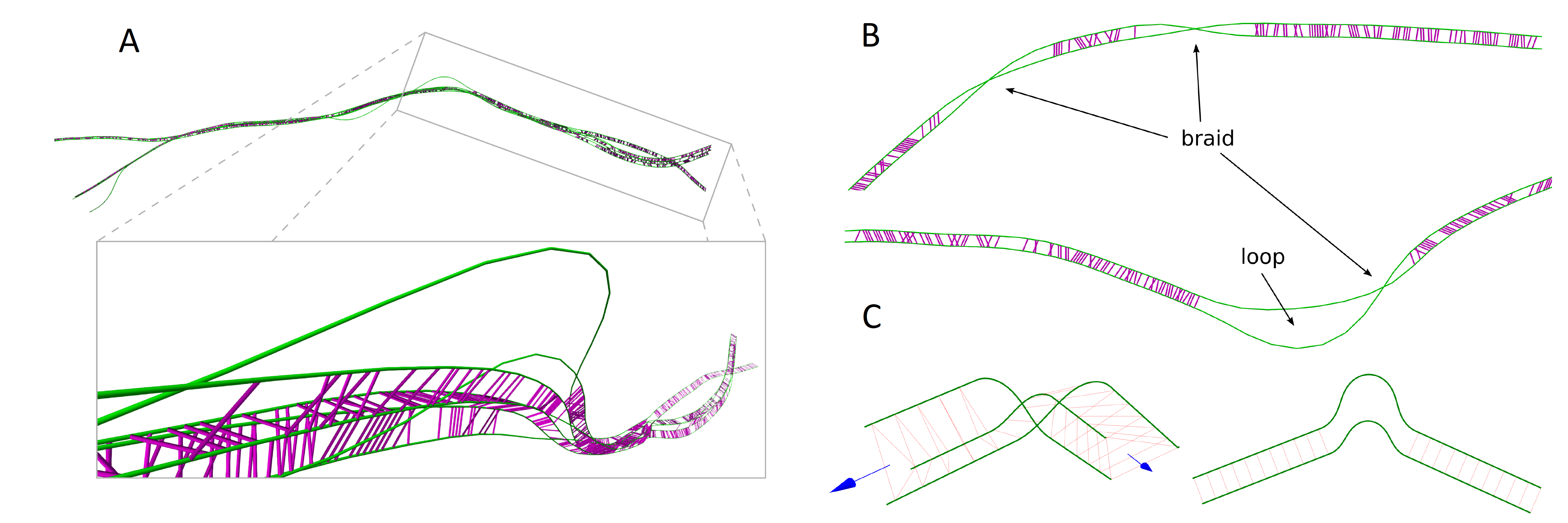}
\caption{Typical bundle shape and defects observed in numerical simulations of the bundle formation 
process, starting from initially straight and parallel filaments without any initial cross links.
(A) Example images show the entire bundle consisting of 7 filaments (green) and approximately 1600 cross 
links (pink). A magnified part is shown in the inset.
(B) Braids and loops observed in simulations with a minimal setup of two filaments (green) and 
transient cross links (pink) in 2D. (C) Schematic of a braid (left) and a loop (right) as described in 
our analytical model.}
  \label{fig:defect_types_in_simulation_and_theory}
\end{figure*}
Our numerical model describes the semiflexible filaments via geometrically exact beam theory and 
includes the random binding and unbinding of cross
links~\cite{Cyron2010,cyron2012,Mueller2015interpolatedcrosslinks}. Bound cross links
are treated as short elastic beams making locally normal connections to the filaments to which 
they are bound.  As a result they act like so-called bundling cross linkers that elastically
constrain the angle between the bound filaments. Such linkers are well-known in F-actin 
networks.  The details of collagen intra-bundle cross linking are more poorly understood. 
In the absence of detailed models for these cross linkers, we 
chose this simple linker model to promote bundling.  
Initially, all filaments were straight and parallel without any cross links.  To form bundles, 
a fixed concentration of cross linkers was added to the finite-temperature (stochastic) simulation.
The interaction of the thermally undulating filaments with transient cross 
linkers lead to rapid bundle self-assembly (see Fig.~S3) with a number of quenched defects. 
Further details of the model and the setup of the computational experiments are provided in the
methods section below and in SI Sec.~2.
%
\subsubsection{Observation and characterization of defects}
A 7-filament bundle is shown in Fig.~\ref{fig:defect_types_in_simulation_and_theory}A 
from a simulation in three dimensions. Its contour deviates quite drastically from the trivial
equilibrium shape of straight and parallel filaments, which are regularly 
cross linked along their entire length. These metastable configurations 
of the bundle with localized bends -- kinks -- persist over long times as compared to the typical 
time scale of the angular fluctuations of the mean local tangent of the bundle. Over still longer times, 
the locations of the kinks move along the bundle, as described below.  A movie
of the bundle dynamics showing the shorter time scale bundle undulations can be found in the electronic SI (see Movie S1).

We observe two distinct classes of defects in the quenched bundles, which are all related to a mismatch 
between amount of filament arc length taken up per fixed unit length of the bundle.  These are (1) 
{\em braids}, i.e., rearrangements of filaments within the bundle, and (2) {\em loops} where one filament 
stores excess length by looping out of the bundle and then reattaching to it.   Both braids (actually
pseudo-braids, as described below) and loops are shown in
Fig.~\ref{fig:defect_types_in_simulation_and_theory}B from a two-dimensional simulation of two 
filaments where the filaments are allowed to cross each other but cannot untwist; this is the smallest
system capable of supporting a loop or a pseudo-braid. The pseudo-braid is a projection of a braid 
onto two dimensions and is the mechanical analog of a true braid in three dimensions, as is discussed 
in Sec.~\ref{sec:Calculations}.  There we show that the two-filament pseudo-braid is energetically 
equivalent to the case of a true braid of three filaments in 3D when the two filaments making up the 
pseudo-braid in 2D have different bending moduli. The simplest system that supports 
true braiding defects is a three-filament bundle in three dimensions, shown in
Fig.~\ref{fig:defect_types_in_simulation_and_theory}C.  The inset of 
Fig.~\ref{fig:defect_types_in_simulation_and_theory}A
shows the typical structure of a loop in a larger bundle.  
There is also  a third type of defect, (3) a {\em dislocation} in which a filament end appears within 
the bundle. This defect was not created in our simulations due to the fact that we started the 
system with equal length filaments whose ends were initially
aligned at one end of the simulation box.  
In the simulations, we concentrate on braids and loops. 
We first analyze the curvature of the bundle center line as well as the kink angles resulting from 
such defects, and then investigate the dynamics of the defects, i.e., how they move along the bundle 
and potentially interact with each other.

\subsubsection{Curvature and kink angles of defected bundles}
Fig.~\ref{fig:kink_images_angles_and_curvature}D shows a typical configuration of the minimal 
bundle setup with two filaments (blue and red).  The bundle centerline (black dashed line) is computed as 
the average of the two filament centerlines, and braids (black dots) 
are detected by the crossings of the filament centerlines.  A movie of the bundle's dynamics can be found in 
the electronic SI (see Movie S2). The curvature of the bundle's centerline as a function of centerline arc length  is 
plotted in Fig.~\ref{fig:kink_images_angles_and_curvature}E, showing 
both the mean (black) and the standard deviation range (red lines) of the curvature computed from 100 
simulation snapshots with a time interval of $1\mathrm{s}$.  Close to the midpoint of the bundle in the  
range of arc lengths~$10 \mathrm{\mu m} < s < 12 \mathrm{\mu m}$, we observe two peaks 
in the curvature that are clearly visible as a double kink in the bundle configuration shown in
Fig.~\ref{fig:kink_images_angles_and_curvature}D.  These can be explained 
by the braid and loop defects there.  The standard deviation of the curvature is increased by about 
one order of magnitude in this defected region, indicating a local increase
in angular fluctuations at this point. This is a direct measure for the decreased effective bending modulus 
of the bundle in these defected, non-cross-linked regions.  Using the relation between the thermal 
fluctuations of the local curvature and the bending 
modulus, we estimate a decrease in the effective bending modulus of about two orders of magnitude.
Apart from the locally decreased bending modulus, such a defect most likely also leads to an anisotropy in 
the bundle's bending mechanics, which breaks another basic assumption of the ideal bundle as a single, 
thick filament.  Similar features in the curvature data are observable for the second braid of this bundle at
approximately $s = 6 \mathrm{\mu m}$ of this bundle.  More examples are found in the other simulation runs. 
Additional results showing the curvature along the bundle at different time points are 
provided in Fig.~S4.

The histogram of measured kink angles over a total of 12 simulations is shown in
Fig.~\ref{fig:kink_images_angles_and_curvature}F.  Here, we applied the same procedure for the angle
measurements as described for the experimentally obtained microscopy images.
The distribution of 72 kink angles for the two-filament bundle has a 
mean of 27 degree, with a standard deviation of 14 degrees and values ranging from 4 to 77 degrees.
The kink angle distribution for larger bundles with 7 filaments demonstrate a trend towards smaller 
angles and a more narrow distribution with 20$\pm$8 degrees (mean $\pm$ standard deviation).
Big bundles with up to 225 filaments will be investigated in more detail below.
\subsubsection{Dynamics and interactions of defects}
\begin{figure}[htpb]
  \centering
  \includegraphics[width=\linewidth]{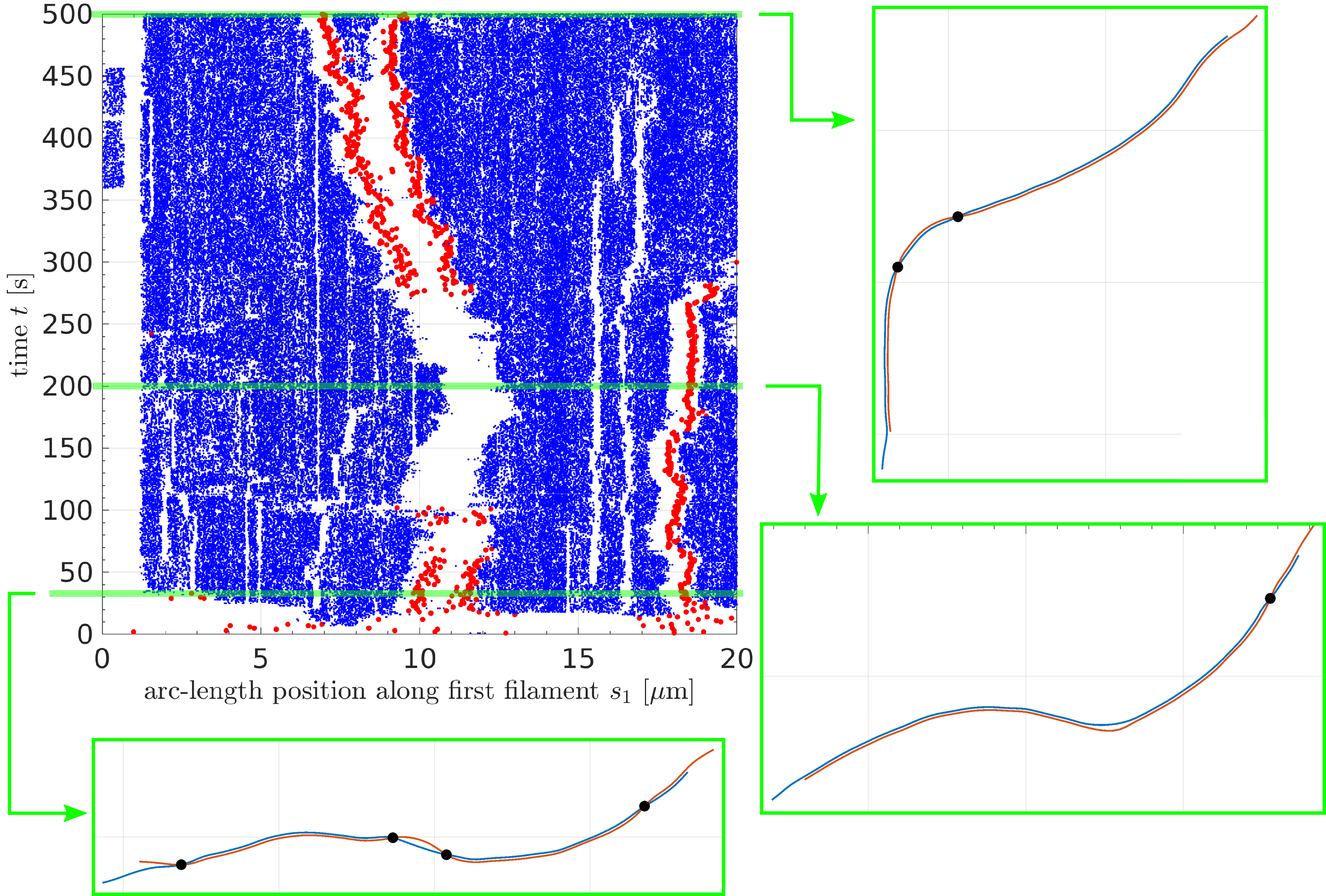}
  \caption{Dynamics and interactions of defects observed in numerical simulation.
           The position of braids (red dots) and cross links (blue dots) along the (first, i.e., blue) filament is tracked over time.
           The inset images (green frames) show the corresponding configuration of the two filaments (blue and red lines) and the braids (black dots) in the bundle at three different time points.}
  \label{fig:dynamics_of_defects}
\end{figure}
We now use our simulations to study dynamics on longer time scales, where we expect to see the motion of 
defects along the bundle and their annealing as the metastable, defected bundle slowly relaxes.  To
facilitate these observations, we need to speed up the motion of the defects by doubling the linker 
unbinding rate in our simulations to $k_\text{off}=6 \mathrm{s}^{-1}$.  At this rate the motion of 
defects is still much slower than the undulatory fluctuations of the bundle, but
now defect motion is moved into a time scale accessible by simulation, which covers 1000 seconds.

Fig.~\ref{fig:dynamics_of_defects} shows an example of how the (defected) configuration of a two-filament 
bundle evolves over time.  We plot the position (measured by arc length)  of braids (red dots) and 
cross links (blue dots) along the bundle horizontally, with time increasing vertically.  The resulting 
red tracks record the world-lines of the braids over a simulated period of 500 seconds.  The white vertical 
scars show cross-linker gaps in the otherwise densely cross-linked bundle. Due to a small off-set 
between the filaments, there is a nearly persistent gap in cross linking at the left end of the bundle 
where one filament stops.  Cross linkers appear in this gap because one filament slid far enough past the 
other to wrap around and briefly cross link to the other one due to the periodic 
boundary conditions of the simulation box.  We observe a pair of braids located near the bundle midpoint first 
emerge after  $\sim 20 \mathrm{s}$ during the initial quench of the bundle. This timescale
for the formation of the quenched bundle is typical and consistent with observation of the initial growth 
of the number of doubly bound cross linkers; that number rapidly increases from zero at the beginning of 
a simulation and plateaus around 20s, indicating the maturation of the defected bundle -- see Fig.~S3 
for further details on bundle self-assembly.
Once the bundle has formed, the two braids in the middle approach each other and appear to annihilate, 
leaving a low-cross-linker density region within the bundle during the time period of 
~$100 \mathrm{s} < t < 270 \mathrm{s}$.  After that time, a new pair of braid defects form. These slowly
separate as more and more cross links are formed between them.  

The single braid close to the bundle's right end diffuses until it  
approaches the far right end of the bundle at $s \approx 19 \mathrm{\mu m}$ and at time 
$t \approx 280 \mathrm{s}$. Here it vanishes by diffusing off the open end.  The filaments simply uncross 
and new cross links are established between the unbraided filaments. Looking more carefully, one may 
observe a similar phenomenon on the left edge of the bundle.  Immediately after the quench there are 
actually four braids on the bundle, as indicated by the picture of the system at timeslice $t=34$s. 
Almost immediately after this time and long before the next bundle configuration image at $t=200{\rm s}$, 
that leftmost braid diffuses off the left end of the bundle. The final state of the bundle at 
$t=500 {\rm s}$ shows a bent configuration where the localized bend near the center of the bundle is 
due to the two interacting braids that remain in the system. For these 
parameter values, the typical lifetime of a defect is 100s of seconds.

\subsubsection{Big bundles}

\begin{figure}[htpb]
  \centering
  \includegraphics[width=\linewidth]{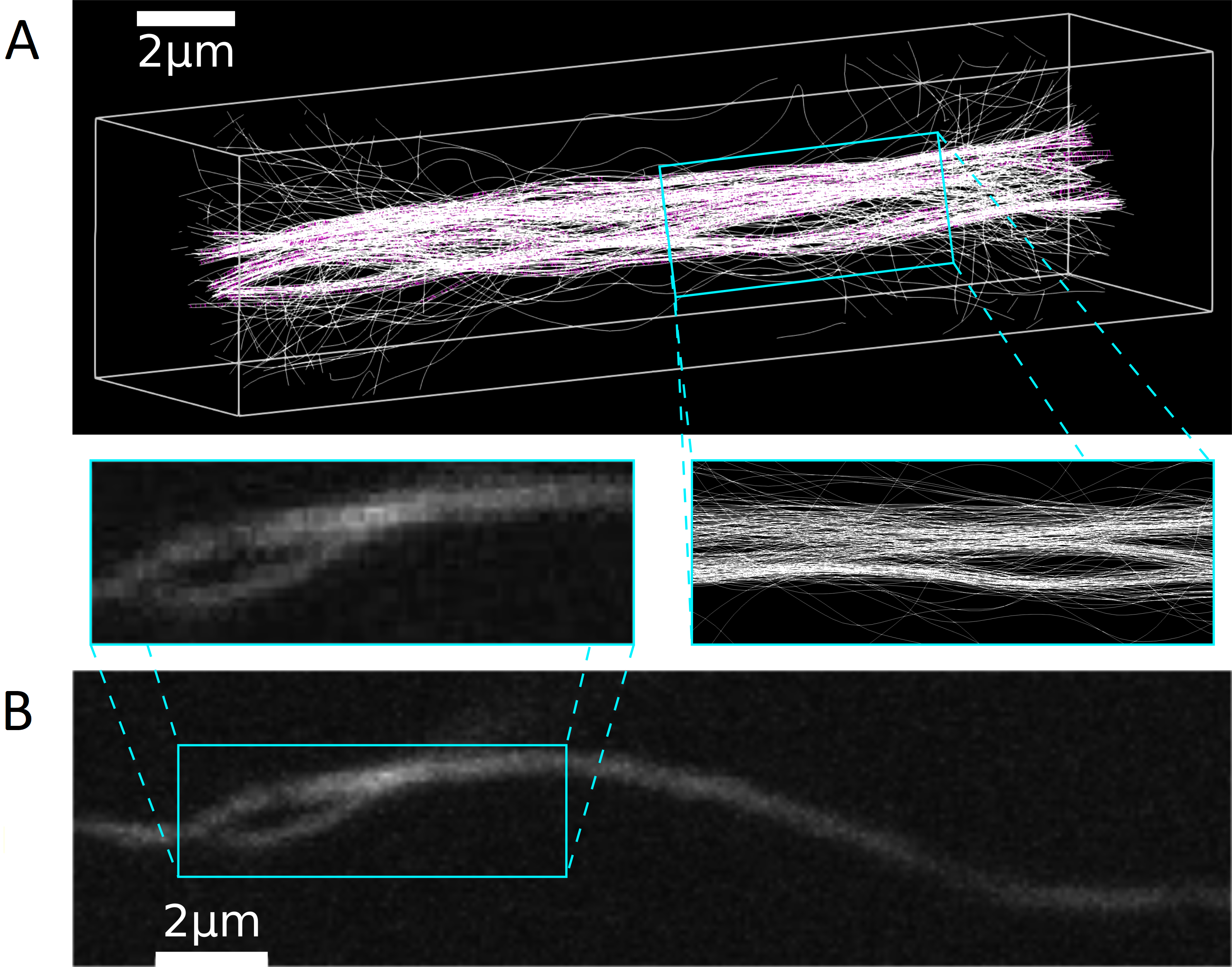}
  \caption{Hole defect observed in simulation and experiment.
           (A) Simulation of a big bundle with 225 filaments (white) and approximately 16000 cross links (pink).
           (B) Fluorescent confocal laser scanning image of a collagen bundle.
           Those parts of both images showing the hole defect are magnified and compared side-by-side in the center panel of this figure.}
  \label{fig:simulation_big_bundle_hole}
\end{figure}
Motivated by the fact that the number of filaments in a biopolymer bundle is likely to vary from 
${\cal O}(1)$ to ${\cal O}(10^3)$, we explored simulations of very big bundles.
Fig.~\ref{fig:simulation_big_bundle_hole}A shows a self-assembled bundle with 
225 filaments (white) and approximately 16000 cross links (pink). To rule out the 
influence of the initial arrangement of filaments in plane perpendicular to the bundle's mean tangent, we ran 
simulations with filament endpoints placed on a square grid in addition to the hexagonal grid.  
We observed no significant differences.

The large bundle's structure is hierarchical; one can identify more tightly bound sub-bundles that form 
loops and braids with each other along the bundle's length.  As observed already in 25-filament bundles, 
its centerline remains rather straight, while the sub-bundles show the characteristic kinks 
observed in the smaller bundles.    One possible explanation for the rather straight form of the 
big bundles is the smaller aspect ratio as compared to the small bundles; in other words, very large bundles 
may well show kinks over longer distances, since such kinks require higher energy 
and thus statistical less probable, defects.  Simulations of big bundles with the same aspect ratio 
as the smaller ones remain computationally prohibitive. We observe in the large bundle a large hole created by 
a sub-bundle loop defect. Its appearance is strikingly similar to
our experimental images of collagen bundles in Sec.~\ref{sec:Experiment}
(Fig.~\ref{fig:simulation_big_bundle_hole}B).
Those parts of the images showing the hole defect are magnified and compared 
side-by-side in the center panel of Fig.~\ref{fig:simulation_big_bundle_hole}.

\subsection{Kinking Theory}
\label{sec:Calculations}

\subsubsection{The model}
\label{subsec:model}
We now examine the energetics of kink formation using a a simple model consisting of a bundle of 
$N$ inextensible, semiflexible filaments connected by cross links. The filaments' elasticity is controlled by 
a single bending modulus $\kappa$. The filaments are arranged so that their mean tangent directions are 
parallel along the $\hat{x}$ axis.  In a cross section normal to this direction (the $yz$ plane) the 
filaments' centers lie on a triangular lattice with a lattice constant equal to the size of the 
cross linkers.  The cross linkers are assumed to locally constrain both the distance between the 
cross-linked filaments and their crossing angle so that the cross links are normal to the filaments to 
which they bind.  We further assume that the cross linking is reversible, i.e.,  they bind and unbind 
from the filament bundle so that the cross linker density within the bundle remains in chemical equilibrium
with a solution of free cross linkers at a fixed chemical potential.  Previous work has 
shown that thermal undulations of the filaments induce Casimir forces between cross
links~\cite{Kachan:13,Kachan:16} and cause the transition between states of free 
filaments and a densely cross-linked bundle to be 
a discontinuous, or first-order phase transition rather than a smooth crossover.  Here we work at 
chemical potentials above this transition so that we may assume dense cross linker 
coverage; hereafter we neglect Casimir interactions and other fluctuation-induced effects.

If all the bundle's filaments have the
same length, the energetic ground state is a straight bundle with as many cross links as possible. 
However, if, at least one of the filament's length differs from those of others, the straight bundle
configuration will necessarily have a defect where a filament's end occurs within
the bundle. That dislocation defect may, in fact, be unstable towards forming a kink in the 
bundle's interior, leading to a  kink in the elastic ground state of the system (we explore
this point in Sec.~\ref{subsec:boundary}). 

When we consider metastable states, there are many more options. If removing a defect in the structure of
cross-linked straight filaments requires uncoupling a large number of cross links, the lifetime of that 
defect may exceed the time of the experiment. We divide such
defects into two groups: defects due to the deviation of the filament from its straight state 
(loop), and the effects due to the permutations of the filaments (braiding).  We study the 
simplest cases of these
effects in Sec.~\ref{subsec:trapped} and Sec.~\ref{subsec:braiding} respectively.

In all these cases, the energy of the bundle can be written as the sum of two terms: the bending 
energies of the ($N$) constituent filaments and energy of their chemical interactions with the cross links
\begin{equation}
E = \sum_{i=1}^N  \int ds \left\{ \frac{\kappa_i}{2} \left(\partial_s \hat{t}_i \right)^2 + \mu \right\},
\label{eq:general-energy}
\end{equation}
where $\hat{t}_i$ is unit tangent vector of filament $i$. The integral is taken over the piece of the 
filament $\ell$ without cross links, which generates
the term $\mu \ell$ equal to the work of unbinding the cross links in this piece, where 
$\mu$ is a linker binding energy per unit length. 
Since we assume that  the cross links completely fix both angle and positions of the filaments, 
the piece of the filament with cross links is straight and parallel to the whole bundle.
We now minimize the bundle's energy subject to boundary conditions that enforce the presence of one or 
more defects.  If a kinked configuration minimizes this energy, we conclude that elasticity theory predicts a kink.  This calculation will also determine the optimal size (length) and 
bending angle of the kink, which we report below. We now perform this minimization for the three different 
types of defects.
\begin{figure}
\centering
\includegraphics[width = 8 cm]{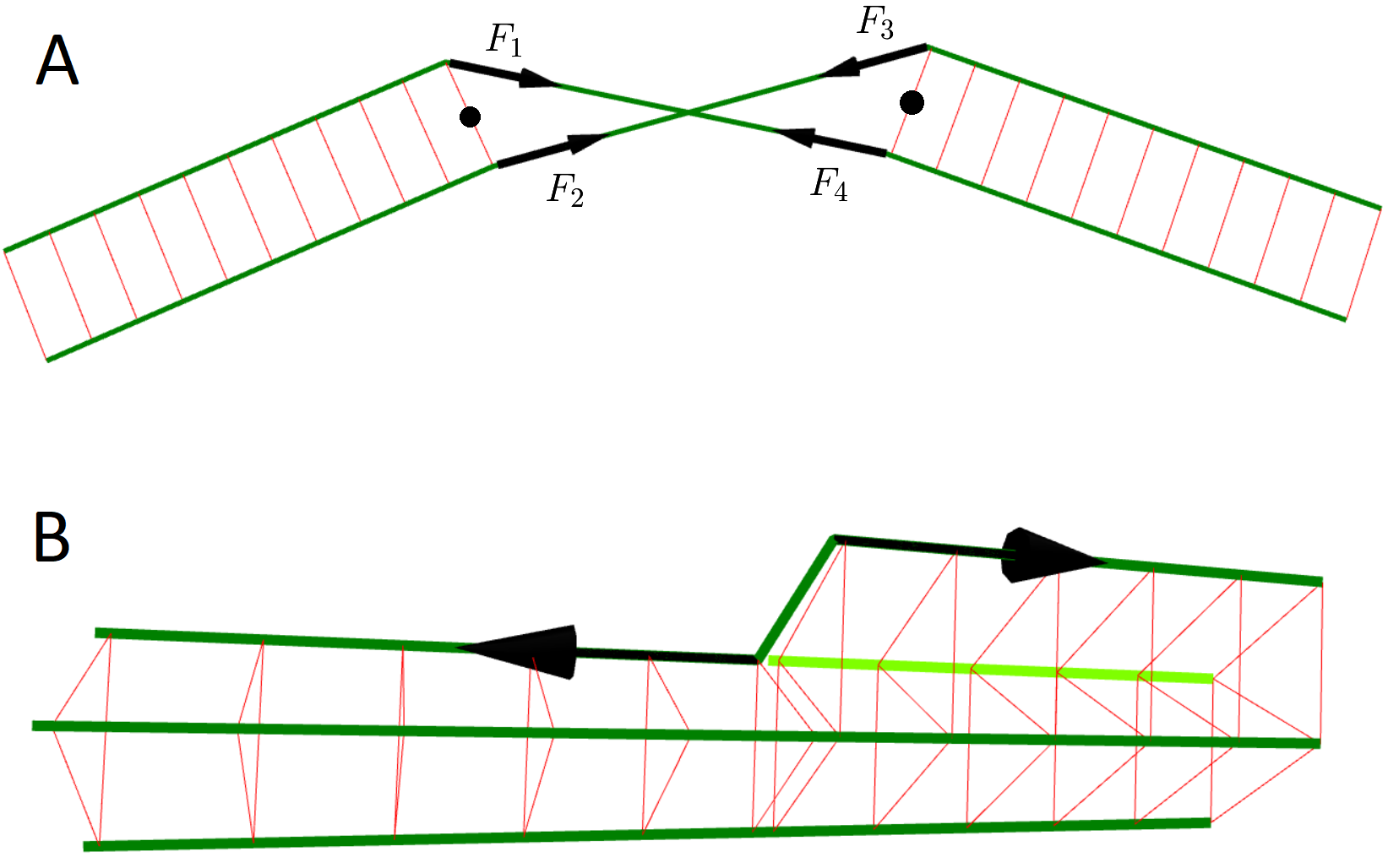}
\caption{(A) Braid in the limit of zero bending. Forces $F_1, F_2, F_3, F_4 $ (black arrows) have 
equal magnitude, but $F_2$ and $F_4$ create a larger torque (relative to the middle of the corresponding cross link, 
black dots). This torque leads to the rotation of the left piece of the bundle counterclockwise and 
right piece clockwise, i.e., increases the angle of the kink. (B) Dislocation in the limit of zero bending. 
The least energy configuration is a straight bundle, with one filament rearranging at the right angle
when filament 4 (chartreuse) stops, immediately taking its place. However, if we increase the bending 
to nonzero, this segment under the right angle tries to straighten, producing repulsive forces (black arrows).
These forces create an uncompensated torque at the left and right part of the bundle, leading to a kink.  }
\label{fig:sketch}
\end{figure}

\subsubsection{Loops}
\label{subsec:trapped}
\begin{figure}
\centering
\includegraphics[width = 7.5 cm]{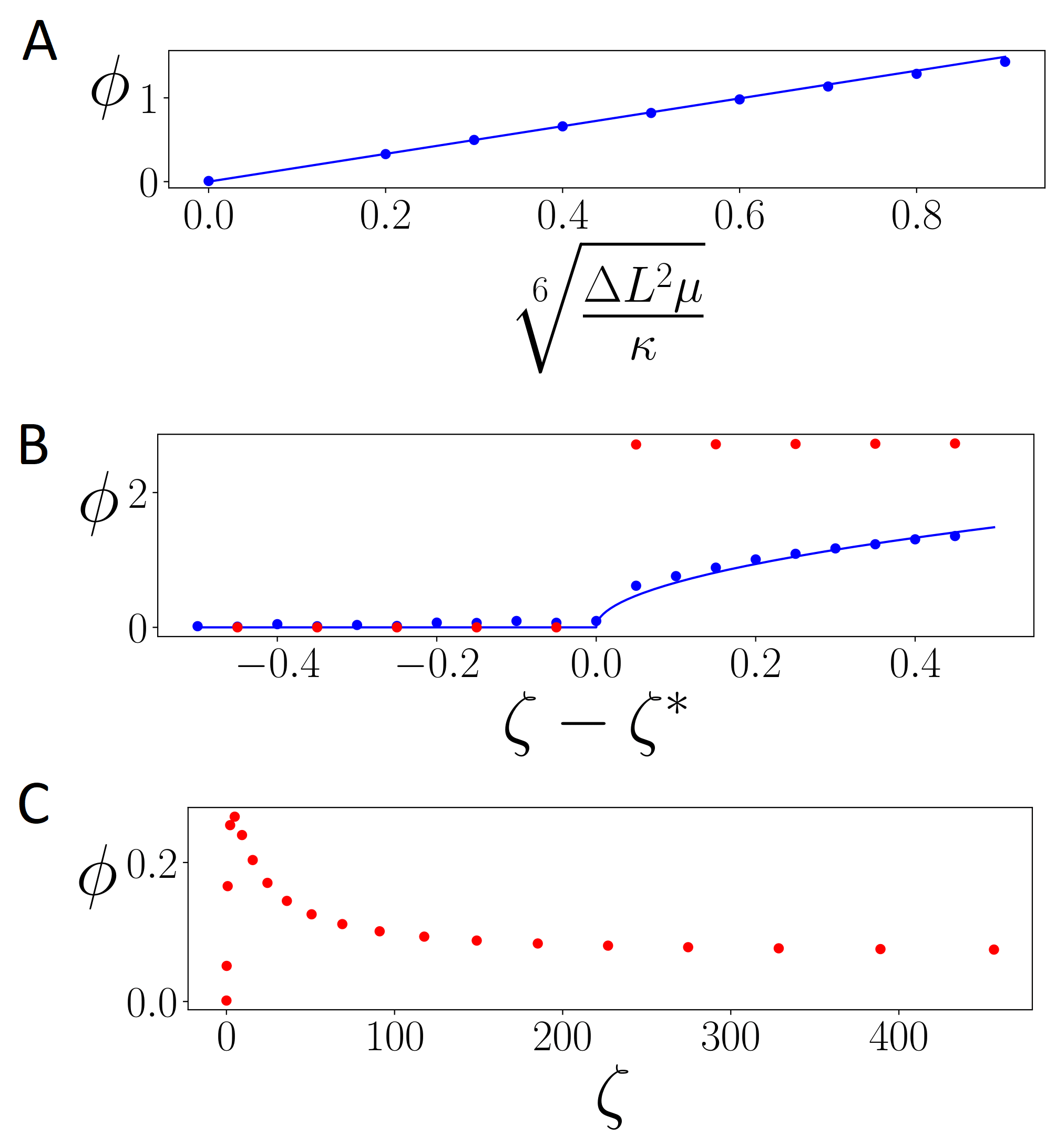}
\caption{Kink angle $\phi$ as a function of a dimensionless parameter for (A) loops, (B) braids and 
(C) dislocations showing both numerical solutions to the energy minimization (circle), and
the analytic predication (solid line). We show two cases $\kappa_1 = \kappa_2$ (blue) and 
$\kappa_1 = 2 \kappa_2$ (red). (A) Numerical results agree with the small angle theoretical prediction 
even up to $\phi \approx \frac{\pi}{2} $. (B)  For the symmetric case $\kappa_1 = \kappa_2$, the 
angle $\phi$  produced by the braid grows as $\sqrt{\zeta - \zeta^*}$ with $\zeta^* = 2$. The 
coefficient of proportionality for the analytical curve was chosen to best fit the data. 
For the asymmetric case, there is a discontinuous (first order) jump in $\phi$ at $\zeta^* \approx 12.25$. 
(C) For dislocations, the angle reaches a maximum at a finite value of $\zeta$ and goes to zero at $\zeta = 0$ and
$\zeta = \infty$.}
\label{fig:angle-loop}
\end{figure}

We start with the simplest case of  a two-filament bundle, forming a loop defect by demanding that the 
filaments have disparate lengths $ L_1 \neq  L_2$ between consecutive cross links. What results is the 
bending of the whole bundle to form a kink -- see Fig.~\ref{fig:defect_types_in_simulation_and_theory}B.  This
approach generalizes to $N$-filament bundles, and can be adapted to large bundles in which two sub-bundles form
a loop. To simplify this calculation, we take  the size of the cross links to be zero. 
Then the boundary conditions for the position of the ends of the filament are integral conditions 
on the tangent vector:
\begin{equation}
\int_{-L_1 / 2}^{L_1 / 2} ds  \hat{t}_1(s)   =  \int_{-L_2 / 2}^{L_2 / 2} ds \hat{t}_2(s),
\end{equation}
and boundary conditions for the tangent vector determine the kink angle $\phi$, which is the 
total bend of the tangent across the structure. We pick a reference frame so that these boundary 
conditions are symmetric:
\begin{equation} 
\label{boundary-condition-angle}
\hat{t}_1 (\pm L_1) = \hat{t}_2 (\pm L_2) = 
\begin{pmatrix}
\cos (\phi/ 2)
\\
\sin (\phi/2)
\end{pmatrix}.
\end{equation}
Minimizing the energy from Eq.~\ref{eq:general-energy} in the limit of small filament 
bending ($\hat t_y \ll 1 $)  we obtain a lengthy self-consistent equation for the angle $\phi$ (see SI Sec. 3A), 
which can be simplified in the case of the equal bending moduli $\kappa_1 = \kappa_2 = \kappa$ to
\begin{equation}
\phi  =   \gamma \left( \Delta L \sqrt{\frac{\mu  }{\kappa} }  \right)^{1/3},
\end{equation}
with the numerical constant $\gamma \approx 0.93$, and $\Delta L = L_2 - L_1 \neq 0$. 
We verified these results by minimizing the energy numerically -- see Fig.~\ref{fig:angle-loop}A.
Loops produce a continuous spectrum of kink angles that grow as the cube root of their length mismatch.  
As expected, an increase in the bending 
modulus suppresses this kink angle, while an increase in the linker binding energy increases it by 
shrinking the extent of the gap in the cross linking. We now turn to braids.

\subsubsection{Braids}
\label{subsec:braiding}
The simplest model of braiding in 3D requires three filaments. Braiding of two filaments in 3D can be 
undone by twisting the bundle about its long axis; it is not topologically protected 
(the relationship between braiding and rotation is discussed in more detail in the SI Sec.~3C).
The minimum energy configuration of three cross-linked filaments with the same length will be a 
right prism with an equilateral triangle as its base. We choose a coordinate system so that the $x$-axis 
lies parallel to the  filaments, filaments 2 and 3 are in $xz$-plane, and filament 1 is above that plane.
To introduce a braid we require filament 1 to go from above to below the $xz$-plane -- see
Fig.~\ref{fig:defect_types_in_simulation_and_theory}C.
This configuration is metastable since we need to decouple all the cross links on one side to get 
to the minimal energy configuration. There is no 
rotation of an end of the bundle that will eliminate the braid.

Since the cross links fix both relative angle and position of the filaments, filament 1 can not be 
connected to the filaments 2 and 3 by the cross links in the defect core, however, filaments 2 and 3 
can remain cross linked. Thus, filaments 2 and 3 behave as one
combined filament 2' with double the bending stiffness, and remains in the $xy$ plane like filament 1. 
The true three-filament braid in 3D is thus energetically equivalent to a two-filament pseudo-braid, 
as introduced in our simulations.

The boundary conditions on the vector $\hat{t}$ are the same as in the previous case,
Eq.~\ref{boundary-condition-angle}, 
but the displacement boundary condition 
differs, incorporating the finite cross linker length $a$, which is necessary for the 
braid to trap excess length.  We find
\begin{equation}
\label{first-condition}
\int_{-L_1 / 2}^{L_1 / 2} ds  \hat{t}_1(s)   = \int_{-L_2 / 2}^{L_2 / 2} ds  \hat{t}_2(s)    
+ 2 a \cos\left(\phi / 2 \right) \hat{y},
\end{equation}
Unlike in looping, we do not fix the filament length mismatch $\Delta L = L_2 - L_1$, but instead 
allow it to vary to relax the braid's energy. It
is conceivable that one may encounter higher-energy braids in which braiding and an excess of 
trapped length (looping) coexist.  We do not study this case here.

We observe that braids should generate local bending, at least in the limit of a sufficiently soft
bending modulus.  The binding free energy (chemical potential difference between free and bound linkers) acts as an 
effective tension on the bundle. 
Setting the bending modulus to zero and fixing the length of the braided region, the solution for the filament contours inside the braid will be
straight lines. In this configuration, linker-induced tension generates a torque that increases the kink angle of kinked configurations -- see 
Fig.~\ref{fig:sketch}A.  To stabilize this angle at a finite value we must including finite bending compliance.  We do so now, turning to the
full calculation.

Calculating the energy of the braid as a function of the kink angle $\phi$, we prove (see SI Sec.~3B) that for 
small values of dimensionless parameters $\zeta_{1,2} = \frac{\mu a^2}{\kappa_{1,2}}$, the energy is minimized 
at $\phi = 0$, i.e., there are no kinks. For large values of $\zeta_{1,2}$, the minimum energy states
are kinked ($\phi>0$).  We examine this transition in more details for the case 
$\kappa_1 = \kappa_2$. We assume symmetric bending, $\alpha_1(s) = -\alpha_2 (-s)$ noting that numerical solutions of the minimization show that the symmetric solution based on this ansatz indeed 
identifies the global energy minimum. We  obtain (see SI):
\begin{equation}
\lambda^2 \sin^2  \left( \phi/2 \right)  = \frac{8 \kappa^2 }{a^4} \left( \frac{\mu a^2}{\kappa}  -  2 \right),
\end{equation}
where we introduced a Lagrange multiplier $\lambda$ to enforce the $\hat{y}$ 
component of Eq.~\ref{first-condition}, which plays the role of the tensile force in 
$\hat{y}$ direction. When the dimensionless control parameter $\zeta = \frac{\mu a^2}{\kappa}$ 
increases to $2$, there is a second order at which the kink angle grows continuously from zero as
$\zeta$ increases. Near the critical point  $\zeta = 2 + \epsilon$, $\phi \sim \sqrt{\epsilon}$. Numerical 
minimization of the energy from Eq.~\ref{eq:general-energy} leads to the same result (see Fig.~\ref{fig:angle-loop}B).

\subsubsection{Dislocations}
\label{subsec:boundary}
The simplest dislocation requires a bundle of four filaments in 3D where one of the four ends within the bundle.
The stable state of four filaments is a right prism formed by a base of  two equilateral triangles 
sharing one edge, as shown in Fig.~\ref{fig:sketch}B. We label these triangles as 1-2-4 and 2-4-3;  
only filaments 1 and 3 are not cross linked to each other. If either
filament 1 or 3 ends within the bundle, the configuration remains stable because the other three form a
stable three-filament prism. But if another filament ends, e.g., filament 4, the remaining filaments 
must deform to recreate a cross section with an equilateral triangle -- see Fig.~\ref{fig:sketch}B.  
Due to cross-linker constraints, the distortion associated with the defect must locally 
remove cross linkers between two of the filaments.  Without
loss of generality, we demand that filaments 2 and 3 remain cross linked. Calculating the energy 
associated with this defect is complicated by the fact that there is no mapping to a 2D version of 
the distortion.  To gain immediate insight, it is helpful to consider momentarily the unphysical 
case of zero filament bending modulus. Then filaments 2 and 3 remain straight, but filament 1 makes two 
right-angle bends at the defect to move to the location of the missing filament 4 and thereby 
maximize cross linking.  If we now reintroduce a finite bending modulus, this localized dislocation will 
spread out along the bundle to decrease bending energy at the expense
of reducing the maximal cross linking shown in the figure.  A force pair is also introduced by 
filament 1's bending (shown in the figure as black arrows) which produce a torque causing the entire 
bundle to kink. We perform numerical minimization of the energy assuming that filaments 2 and 3, being 
cross linked everywhere, form a ribbon that can bend in the direction perpendicular to its plane with 
bending modulus $\kappa_{ribbon} = 2 \kappa$, but is absolutely rigid in the direction parallel to its 
plane. The results are consistent with the qualitative study (see Fig.~\ref{fig:angle-loop}C) -- the 
maximum of the kink angle is observed at a finite value of parameter $\zeta$, while zero and infinite values
lead to zero kink angle.

\subsection{Defect dynamics}
\label{subsec:lifetime}
Over times significantly longer than those associated with the undulations of the bundle, defects can move along 
the bundle and interact.  These dynamics require multiple cross linker binding/unbinding events.  
As a result of these events, defects move diffusively and may eventually fall off the ends of the bundle.  
In the case of dislocations and braids, defects may combine or annihilate.  For the latter type, these
interactions are controlled by the structure of the braid group.  

Consider two braids -- a braid/anti-braid pair -- separated by $N$ cross links. Since these defects 
would annihilate if the intervening cross links were removed, we may expect this pair might vanish if 
their separation becomes sufficiently small.  The braids are motile with a diffusion constant set by the 
linker detachment rate  $k_{\rm off}$ and do not strongly interact when separated by 
lengths greater than the defect core size.  The probability density $p(n,t)$ of there being 
$n$ cross links separating the two defects at $t$ then obeys a diffusion equation 
\begin{equation}
\frac{\partial p(n,t)}{\partial t} = 2 k_{\rm off} \frac{\partial^2 p(n,t)}{\partial n^2}
\end{equation}
Using a well-known result for the first mean passage time~\cite{vankampen2007spp}, the mean 
lifetime of this braid/anti-braid pair is
\begin{equation}
T = \frac{3 N^2}{ k_{\rm off}}.
\label{mean-passage}
\end{equation}
For the simulations presented in Fig.~\ref{fig:dynamics_of_defects} we have $k_{\rm off} = 6 \mbox{s}^{-1} $, 
$N \approx 15 - 35$. Then Eq.~\ref{mean-passage} predicts $T \approx 100 - 500\mbox{s}$, while 
in Fig.~\ref{fig:dynamics_of_defects} we obtain $T  \approx 200\mbox{s}$, within the predicted range.

For a three-filament bundle the dynamics of $N$ braids is equivalent to the diffusion of 
$N$ particles (braids) of three types, which are randomly distributed after a quench.  The braid
group (see SI Sec.~3C) requires that a particle of one type can annihilate only with particles of one other type.
If particles encounter each other and cannot annihilate, we assume they stick, since, by merging their 
defected regions, the net number of cross linkers on the bundle increases. 
Using these dynamical rules, we studied the annealing of braided bundles using Monte-Carlo simulations -- results are shown in the SI.

Simple combinatorics shows that annihilation events are less common than braid combination (sticking) 
since the former requires braid/anti-braid adjacency.   Since the number of different 
braid group operators grows linearly with the number of filaments in the bundle, the probability 
for braid/anti-braid adjacency decreases with increasing braid size. When considering large bundles we can
neglect annihilation.  Doing so and using a mean-field approximation, we let  $\rho(x,t)$ be the 
braid density, implying that the average distance between neighboring braids is  $ \frac{1}{\rho}$. The 
time to halve the number of braids will then be $t_{1/2} \propto  \frac{ 1}{ \rho^2 k_{\rm off}}$ according to Eq.~\ref{mean-passage}. The same logic implies that the continuous rate of decrease of the 
braid density will obey
\begin{equation}
\label{braid-density-decay}
\frac{d \rho}{d t} = - \alpha \rho^3
\end{equation}
where  $\alpha$ is a phenomenological parameter accounting for the probability of braids combining upon
close approach. Solving Eq.~\ref{braid-density-decay}, we find  $\rho \propto t^{-1/2} $, which is a 
general result for sticky (or annihilating)  
random walkers~\cite{Cardy:96}.  The predictions of this mean-field model are consistent with 
our Monte Carlo simulations and with the 
Brownian dynamics simulations of the full bundle model - see Fig.~S5. 

We briefly mention the dynamics of loop and dislocation defects.  
Complete annihilation of loop defects is highly unlikely as it would require the amount of trapped 
length in the two loops to match. We expect loop defects to diffuse along the bundle and, in larger bundles, 
to pass through each other.  Dislocations should also diffuse by a type of 
reptative motion (as in polymer melts) in which the filament end detaches within the bundle, 
forms a loop and reattaches.  Thus, dislocations in an
otherwise ordered bundle should retract towards the bundle edge with more filaments in it.  
After loops are formed the dislocation should 
perform a biased random walk due to the fact that the energy of loop defects will suppress further 
retractions of the dislocation core towards the bundle's end.

\section*{Discussion}
\label{Discussion}

Biopolymer filament bundles are kinked in spite of the fact that the elastic ground state of their 
constituent filaments is straight, as clearly seen in our experiments on collagen bundles.  In this article, 
we quantified these kinks and proposed that their existence can be attributed to defects 
quenched into the bundles during cross linking.   
These defects come in three classes: loops, braids, and dislocations.  This proposal is supported by both 
analytic calculations of the energy-minimizing contour of bundles containing these defects and by 
finite-element Brownian dynamics simulations of the quenched bundles of two to two hundred 
filaments.  The mechanical connection between these defects and kinks (high curvature regions) of the 
bundle is straight forward -- defects generate a local distortion of the filaments driven by cross linking. 
The entire bundle may bend producing a kink in order to compensate for that distortion. This mechanism 
is reminiscent of the relaxation of a flexible hexatic membrane in the vicinity 
of a disclination~\cite{park1996topological}.  There a topological defect relaxes local strain via a 
puckering of the membrane that produces long-ranged Gaussian curvature. Here the distortion of the bundle 
may be entirely localized in a sharp bend.  

In our experiments we found that 
4\% of the observed collagen bundles had one or more kinks and that these kink angles had a mean 
of 26 degrees, but were quite varied ranging up 74 degrees in the sample of 74 kinks studied. When we consider that loop defects can produce a 
continuous distribution of kink angles, it seems natural to suppose that this defect is the predominant cause of kinking. The number of observed kinks is likely an underestimate of the real system, due to the limitations of our imaging that shows only those bundles lying in the imaging plane.  Only kinks oriented so that the bundle bends within the imaging plane are observable.  

The kinks associated with both braids and dislocations are expected to be narrowly distributed at 
angles set by the number of filaments in the bundle since these defects produce fixed kink angles that 
depend only on that number, the cross-linker binding energy, and bending moduli
of the filaments.  For a fixed number of filaments, both dislocations and braids produce kink angles that 
depend on only a single dimensionless number 
$\zeta = \mu a^2/\kappa$. In the case of braids, the kink angle grows  from zero at a 
critical value of $\zeta$, which
depends on the number of filaments but is roughly of order unity.  Looking at stiff F-actin 
cross linkers like $\alpha$-actinin, we 
find that $\zeta \sim 0.1$; it is too small for braids to generate kinks. 
We do, however, expect braids to be associated with kinks in 
softer filament systems such as DNA condensed by polyvalent counter 
ions~\cite{Manning:06,Raspaud:98} or cross-linked intermediate
filaments~\cite{Charrier:16,HAIMOV:2020} where $\zeta \sim 10-100$. Currently, our understanding 
of collagen bundle cross linking is less precise; our estimate in this system is that 
$\zeta \sim 1$ (see SI Sec.~1D).
This suggests that loops certainly should produce kinks, but that braids are also potentially kink-generating
defects since our estimate for $\zeta$ is near the threshold where such braid-induced kinking should occur. 
Of course, even if braids do not produce kinks, we expect them to be present and to produce high flexible 
``hinges'' in the bundle. 
Dislocations always generate kinks, but the kink angle is appreciable only when $\zeta \sim 1$. 
We surmise that dislocations may also be responsible for some of the experimentally observed 
kinks in the collagen bundles.

Another argument for loop controlled defects in collagen is a presence of z-shaped double kinks (see Fig.~S6 for the examples), which can be attributed to slippage between two filaments in a bundle such that they produce a pair of loops.  The lengths stored in this pair are such that, after the two loops, the filaments once have no length mismatch.

The life time of these defects appears to be significantly larger than the 
characteristic time of thermal undulations of the filaments and longer than the typical 
observation time in experiment. 
This is supported by the experimental data, where kink annihilation or diffusion to the 
ends is never observed.  When we study kink dynamics via simulation on the timescales significantly 
longer than those covered by experiment, we observe their diffusion, 
sticking, and annihilation, which one expects from the theory.  Specifically for braids, 
we find that their motion is consistent with  
particles diffusing in 1D with interactions obeying the rules of the braid group. We speculate that 
bundles under compression may relieve stress by the pair production of 
braid/anti-briad pairs in a manner resembling the Schwinger
effect~\cite{schwinger1951gauge,schwinger1954theory}. 

Examining Fig.~\ref{fig:simulation_big_bundle_hole} leads us to speculate that very 
large bundles of many filaments might be considered to be smaller bundles composed of more weakly 
bound sub-bundles, which are themselves composed of the original filaments.  
If we may consider this hierarchical approach, we can 
replace $a$ in $\zeta$'s by the sub-bundle 
radius and write the bending modulus in terms of that radius as well using 
$\kappa \sim E a^4$, where $E$ is the Young's modulus of the 
material (typically in the 1GPa range for proteins).  In that case, we see 
that $\zeta \sim (\mu/E)a^{-2}$, so that as the radius of the
sub-bundles increases, $\zeta$ decreases rapidly.  As a result, we expect 
that kinks in larger bundles will be dominated by loop defects 
regardless of the value of $\zeta$ for the original filament system.  

We note that defects rather generally produce weak links in the bundle where, 
due to the absence of cross linking, the effective bending 
modulus of the bundle is reduced by at least an order of magnitude.  
This suggests that the collective mechanics of a 
rapidly quenched bundle network might be dominated by these defects, which 
introduce a set of soft joints into the otherwise quite stiff 
bundles. As a result, rapidly quenched bundle network may be anomalously 
compliant as compared to their annealed state.  It is 
interesting to note that these defects provide soft hinges in the network 
(rather than universal joints) and that there may well be 
many more such soft hinges than there are kinks, since not all defects generate 
kinks, but all disrupt the local cross linking.  Currently,
there are no kinetic theories of bundling that allow us to estimate the number of 
such soft hinges in a network of filament bundles and 
then attempt to predict the mechanics of the defected network.  
This remains an interesting direction for future studies.

\matmethods{ \label{sec:Model}
\subsection*{Experiments}
Type I Bovine pepsin extracted collagen (PureCol 5005-100ML lot 7503, Advanced BioMatrix) was 
reconstituted according to Doyle, 2016 \cite{Doyle:16}. Reconstituted collagen solution was diluted to 
0.2mg/mL with PBS and was incubated at 37$^\circ$C over night. The collagen was fluorescently labelled 
(Atto 488 NHS ester 41698-1MG-F lot BCBW8038, Sigma-Aldrich) and then imaged with Olympus Fluoview1200 
laser scanning confocal microscope using a 60x 1.45NA oil immersion objective. To construct a trace of the 
bundle, Matlab was used to determine the position of bundle in each row of the image defined as the mean of 
the Gaussian fit of the pixel intensity across each row. A cubic spline is used to estimate the curvature 
along the bundle. The kink angles were measured using imageJ.
Further details can be found in Sec.~1 of the SI.

\subsection*{Simulations}

In our numerical model, the individual semiflexible filaments are described via nonlinear, 
geometrically exact, 3D Simo-Reissner beam theory~\cite{reissner1981,simo1985} and discretized in 
space by suitable finite element formulations~\cite{Jelenic1999,Meier2017b}.
Their Brownian dynamics is modeled by including random thermal forces and viscous drag forces along the 
filament~\cite{Cyron2010,cyron2012}.
We apply an implicit Euler scheme to discretize in time, which allows us to use relatively large time 
step sizes~\cite{Cyron2010,cyron2012}.  Cross links are modeled as additional, short beam elements 
between distinct binding sites on two filaments, which bind and unbind randomly based on given 
reaction rates and binding criteria~\cite{Mueller2015interpolatedcrosslinks}.
In particular, the latter include a preferred distance between binding sites and a preferred angle 
between filament axes that need to be met such that a linker molecule switches from the free to the 
singly bound state or from the singly to the doubly bound state.
Altogether, this finite element Brownian dynamics model turns out to be a highly efficient numerical 
framework, which enables large-scale simulations with hundreds of filaments over hundreds of seconds and 
has been used in several previous
studies~\cite{Cyron2013phasediagram,Mueller2014rheology,Mueller2015interpolatedcrosslinks,Maier2015,Kachan2016bundlingcasimir,Slepukhin2019}.
We used the existing C++ implementation in our in-house research code BACI~\cite{BACI2020}, 
which is a parallel, multi-physics software framework.  In addition, we used self-written 
Matlab~\cite{MATLAB2017b} scripts for the data analysis and Paraview~\cite{Paraview580} for the 
visualization of the system.  Further details about the numerical model including all parameter values 
and the detailed setup of the computational experiments can be found in Sec.~2 of the SI.

\subsection*{Supplemented Information} 

SI can be found at \url{https://drive.google.com/drive/folders/1JF9Dq6kLFWOoAiWDgJwSScRwbZLWiqaz?usp=sharing}
}

\showmatmethods{} 

\acknow{AJL and VMS acknowledge partial support from NSF-DMR-1709785
VMS acknowledges support from the Peccei scholarship and the Bhaumik Institute 
Gradaute Fellowship. WAW and MJG acknowledge partial support from BaCaTeC.
EB and QH acknowledge support from the US Air Force Office of Scientific Research FA9550-17-1-0193 and the Office of the President of the University of California.}

\showacknow{} 

\bibliography{bib_kei,bib_bundles}
\end{document}